\documentclass[runningheads]{llncs}
\usepackage[T1]{fontenc}
\usepackage{amsmath,amssymb}
\usepackage{graphicx}
\usepackage[utf8]{inputenc}

\usepackage[linesnumbered,ruled]{algorithm2e}
\usepackage{graphicx,xcolor,array}
\usepackage{array}
\usepackage{enumitem}
\usepackage{makecell}
\usepackage{pifont}
\usepackage{threeparttable}

\newtheorem{researchproblem}{\bf Research Problem}
\newcommand{\cmark}{\ding{51}}
\newcommand{\xmark}{\ding{55}}
\begin{document}
\title{SoK: Decentralized Randomness Beacon Protocols}

\author{Mayank Raikwar\orcidID{0000-0002-5479-5748} \and
Danilo Gligoroski\orcidID{0000-0002-7078-6139}}
\institute{Norwegian University of Science and Technology (NTNU)
Trondheim, Norway
\email{\{mayank.raikwar,danilog\}@ntnu.no}}
%
%
%
\maketitle              
\begin{abstract}

The scientific interest in the area of Decentralized Randomness Beacon (DRB) protocols has been thriving recently. Partially that interest is due to the success of the disruptive technologies introduced by modern cryptography, such as cryptocurrencies, blockchain technologies, and decentralized finances, where there is an enormous need for a public, reliable, trusted, verifiable, and distributed source of randomness. On the other hand, recent advancements in the development of new cryptographic primitives brought a huge interest in constructing a plethora of DRB protocols differing in design and underlying primitives.

To the best of our knowledge, no systematic and comprehensive work systematizes and analyzes the existing DRB protocols. Therefore, we present a Systematization of Knowledge (SoK) intending to structure the multi-faced body of research on DRB protocols. In this SoK, we delineate the DRB protocols along the following axes: their underlying primitive, properties, and security. This SoK tries to fill that gap by providing basic standard definitions and requirements for DRB protocols, such as Unpredictability, Bias-resistance, Availability (or Liveness), and Public Verifiability. We classify DRB protocols according to the nature of interactivity among protocol participants. We also highlight the most significant features of DRB protocols such as scalability, complexity, and performance along with a brief discussion on its improvement. We present future research directions along with a few interesting research problems.

\end{abstract}

\section{Introduction}

 Public digital randomness is an essential building component for a large spectrum of applications and protocols. For example, a reliable source of continuous randomness providing high entropy, also known as \textit{random beacon}, is crucial for many security applications. A notion of \textit{coin tossing protocol}~\cite{blum1983coin} was proposed by Blum in 1983 that addressed the question of generating the trustworthy random value in a network of mutually distrustful participants. Further, Rabin~\cite{rabin1983transaction} formalized the notion of the random beacon. Since then, randomness generation has been advanced significantly due to the underlying modern cryptography.
 
 
 Lately, coin tossing protocols became more appealing in Proof-of-Work (PoW) or Proof-of-Stake (PoS)~\cite{gilad2017algorand} consensus. Random beacon has a range of applications that includes cryptographic parameter generation~\cite{lenstra2015random}, design of byzantine fault tolerant (BFT) protocols~\cite{cachin2005random,gilad2017algorand}, privacy-preserving message services~\cite{goel2003herbivore}, e-voting protocols~\cite{adida2008helios}, online gaming~\cite{bonneau2015bitcoin}, publicly auditable selections~\cite{bonneau2015bitcoin}, anonymous browsing~\cite{goulet2015random}, sharded blockchains~\cite{croman2016scaling} and smart contracts~\cite{kosba2016hawk}.
 
 
 Due to the applicability of shared randomness in a variety of applications, a rich body of literature has emerged that proposes many DRB protocols differing in their designs and underlying cryptography. Nevertheless, the system models and design challenges in these DRB protocols are highly disparate. Therefore, to address these challenges and to provide a general definition of a DRB protocol, we present a Systematization of Knowledge (SoK). The purpose of this SoK is to provide a systematic overview of existing DRB protocols that can help researchers and practitioners to find suitable solutions for randomness generation. 

\paragraph{\textbf{Background.}}
An easy approach to achieve continuous randomness is through a single node or a trusted third party such as NIST~\cite{kelsey2019reference}, Random.org~\cite{haahr2010random} or Oraclize.it~\cite{Oraclize}. The NIST beacon continuously outputs hardware-generated random values from a quantum-mechanical effect. Since these beacon services are centralized, they can be unreliable. Moreover, in the past, they suffered a significant public trust deterioration after the revealed backdoor in the NSA-designed Dual elliptic curve pseudorandom number generator~\cite{shumow2007possibility}. Due to these problems, these centralized beacon services are undesirable for secure applications. 

As a consequence, \textit{Decentralized Randomness Beacon} (DRB) protocols were proposed and constructed where trust is distributed across multiple nodes that jointly generate random values at a regular interval. More concretely, a consortium of organizations launched a distributed publicly verifiable randomness beacon that periodically provides unbiasable and unpredictable random outputs. The deployment is known as League of Entropy (LoE))~\cite{LoE}
that aims to provide collaborative governance for protection of its random beacon. 
The consortium believes that their beacon can become a fundamental service on the internet.

DRB protocols can be constructed by employing different cryptographic primitives e.g. Publicly Verifiable Secret Sharing schemes (PVSS)~\cite{kiayias2017ouroboros,syta2017scalable,schindler2020hydrand,cascudo2017scrape,cascudo2020albatross,bhat2020randpiper,das2021spurt}, Threshold Crypto-Systems~\cite{cachin2005random,hanke2018dfinity,Drand,cherniaeva2019homomorphic,nguyen2019scalable}, Verifiable Random Functions (VRF)~\cite{david2018ouroboros,gilad2017algorand,galindo2020fully,wang2020randchain,Corestar,DAOBet}, Verifiable Delay Functions (VDF)~\cite{lenstra2015random,drake2018minimal,ephraim2020continuous,schindler2021randrunner,han2020randchain}. Randomness can also be extracted from external data sources such as~\cite{bentov2016bitcoin,clark2010use,bonneau2015bitcoin} or from the blockchain schemes having their own random beacon~\cite{kiayias2017ouroboros,galindo2020fully,hanke2018dfinity}. These DRB protocols are not equally-suited in all applications or use-cases due to the diversity in their designs, characteristics, and underlying assumptions. 


DRB protocols differ significantly due to their underlying techniques. A DRB protocol should have a list of desirable beacon properties along with low communication complexity, low computational cost, and low trust requirement (e.g., setup assumptions). Additionally, the DRB protocol should be efficient in practical settings. Therefore, despite having many constructions of DRB protocols, a few problems such as scalability, trust, and network assumptions need to be addressed to construct a desirable DRB protocol for practical applications.

Motivated by the above, the contributions of this SoK are as follows:

\begin{enumerate}
    \item We provide a formal definition of a Decentralized Randomness Beacon (DRB) with a brief description of its security properties (Sect.~\ref{Sec:DRB}).
    \item We present a classification of DRB protocols in Interactive and Non-interactive DRB protocols and we describe these protocols in detail  (Sect.~\ref{Sec:DRB-classification}).
    \item We give a brief discussion on several crucial issues related to DRB protocols, including complexity, scalability, and assumptions. We also identify a few efficient building components to construct efficient DRB protocols (Sect.~\ref{Sec:Discussion}).
\end{enumerate}


\section{Decentralized Randomness Beacon (DRB)} \label{Sec:DRB}
 A DRB allows a group of participants to collaboratively produce random values without the need of a central party. A DRB consists of $n$ participants~\footnote{We use node and participant interchangeably in protocols throughout the paper.} $\mathcal{P} = (P_1, P_2, \ldots, P_n)$. These participants are connected in a distributed manner. Without loss of generality, we assume that a DRB protocol works in rounds and maintains a beacon state $st$ for each round. For every round $e \in \{1, 2, \ldots\}$, given the current state $st_{e-1}$, the DRB protocol collectively produces a random output $v_e$; the state $st_0$ is jointly computed and agreed from the protocol participants during the bootstrapping of the DRB protocol. Following, we present a formalization of DRB and we formally define the required security properties of a DRB. Additionally, we define a secure DRB protocol in Appendix~\ref{appendix:secure-DRB}.

\begin{definition}{(Decentralized Randomness Beacon (DRB))} \label{Def:DRB}
A DRB on a set of participants $\mathcal{P} = (P_1, \ldots, P_n)$ is defined as a tuple $\mathcal{B}$ of polynomial algorithms:\\
$\mathcal{B} = \mathsf{(Setup, LocalRand, GlobalRand, VerifyRand, UpdateSt)}$
\begin{itemize}[leftmargin=*]
    \item $\mathsf{Setup}(1^{\lambda},n)$: Given input security parameter $\lambda$, and $n$ participants, it generates public parameter $pp$ and keypair for each participant $(pk_i,sk_i)$. All participants agree on public parameter $pp$ and $\{pk_i\}$.
    \item $\mathsf{LocalRand}(st_{e-1},pp,sk_i,s_{e,i})$: Given input state $st_{e-1}$ from round $e-1$, public parameter $pp$, and input seed $s_{e,i}$, a participant $P_i$ computes a local output value $v_{e,i}$ with a proof ${\pi}_{e,i}$ using $sk_i$ and $s_{e,i}$ for round $e$. Output $(i,v_{e,i},{\pi}_{e,i})$.
    \item $\mathsf{GlobalRand}(st_{e-1},pp,\mathcal{S} = \{(i,v_{e,i},{\pi}_{e,i})\},m)$: Given input state $st_{e-1}$, public parameter $pp$, a set $\mathcal{S}$ of local output values from $|\mathcal{S}|$ participants, if $|\mathcal{S}| \geq m$, where $m$ is the minimum number of required local output values, the algorithm computes the beacon output $v_e$ for round $e$ by executing a function $f$ on $\{v_{e,i}\}$ from set $\mathcal{S}$. It also computes proof of correctness ${\pi}_e$ using $\{ {\pi}_{e,i} \}$ from set $\mathcal{S}$. Output $(v_e,{\pi}_e)$ or $\perp$.
    \item $\mathsf{VerifyRand}(st_{e-1},pp,v_e,{\pi}_e)$: Given input state $st_{e-1}$, public parameter $pp$, a beacon output $v_e$, and a proof ${\pi}_e$, the algorithm verifies the beacon value $v_e$ and the corresponding proof ${\pi}_e$. Output $\mathsf{0}$ or $\mathsf{1}$.
    \item $\mathsf{UpdateSt}(st_{e-1},pp,v_e,{\pi}_e)$: Given input state $st_{e-1}$, public parameter $pp$, a beacon output $v_e$, and a proof ${\pi}_e$ generated at the round $e$, the algorithm updates the state from $st_{e-1}$ to $st_e$ for round $e$. Output $st_e$ or $\perp$.
\end{itemize}

 \end{definition}

The security properties of a DRB corresponds to: \textit{Unpredictability}: An adversary should not be able to predict (precompute) future beacon outcomes; \textit{Bias-Resistance}: A single participant or a colluding adversary cannot bias the future beacon values; \textit{Availability (or Liveness)}: A single participant or a colluding adversary can not prevent the generation of the new beacon value; \textit{Public Verifiability}: Any third party can verify the correctness of the new beacon value. \textbf{Note:} \textit{We use DRB protocols and DRBs interchangeably throughout the paper}.

These formal security guarantees of a DRB protocol are evolved during the time. Initial proposals lack the formal definitions and mathematical proofs of their DRB protocols. Nevertheless, the recent proposals put an emphasis on the security of their protocols. These protocols define and prove the security properties of their DRB using the mathematical properties of the underlying cryptographic primitives. Due to different designs, setting up a formal provability framework for DRBs should define the least common security requirements, therefore, we formulate the desiderata of a DRB protocol as follows where $\lambda$ is a security parameter and $\mathsf{negl}(\lambda)$ is a negligible function of $\lambda$.

\begin{definition}{(Unpredictability)} 
Let $\mathcal{A}(v_1, \ldots,v_{e}, st_e)$ be a probabilistic polynomial time algorithm that receives the values $v_1, \ldots,v_{e}$ and the current state $st_e$ as the input values. Let $\mathcal{A}$ outputs a value $v_{e+f}$ for any value (future rounds) $f \geq 2$, and for all rounds $e \geq 1$. Then 
\begin{equation}
    \Pr [\mathcal{A}(v_1, \ldots,v_{e},st_e) = v_{e+f}] \leq \mathsf{negl}(\lambda)
\end{equation}
\end{definition}

\begin{definition}(Bias-Resistance)
Let $\mathsf{bit}_{i}(v_e) $ denotes the $i$-th bit in the binary representation of $v_e$, let $b = |v_e|$ is the number of bits of $v_e$, and let $\mathcal{A}_i(v_1, \ldots,v_{e-1}, \\st_{e-1})$ for $i=1,\ldots,b$, be $b$ probabilistic polynomial-time algorithms that receive the values $v_1, \ldots,v_{e-1}$ and the current state $st_{e-1}$ as input and output one bit: $0$ or $1$. Then for every round $e \geq 1$, every  $\mathcal{A}_i(\ )$ and for all $i=1,\ldots,b$
\begin{align}
    \Pr [\mathsf{bit}_{i}(v_e) & = \mathcal{A}_i(v_1, \ldots,v_{e-1}, st_{e-1})] \leq \frac{1}{2} + \mathsf{negl}(\lambda)\\ 
    \Pr [\mathsf{bit}_{i}(v_e) & = 0] \leq \frac{1}{2} + \mathsf{negl}(\lambda)
\end{align}
\end{definition}
More concretely, we say that a DRB protocol is Bias-resistant if predicting any single bit of the random beacon output $v_e$ has only a non-negligible advantage over the trivial guessing strategy that has a probability of 1/2.

\begin{definition}(Availability)
Let $\mathcal{A}$ be an adversary controlling a fraction of participants and ${\mathcal{P}}^h \subseteq{} \mathcal{P}$ be a set of honest participants in the DRB protocol. Given $v_e, {\pi}_e, pp$ and $st_{e-1}$, for every round $e \geq 1$ and for every participant $P_i \in {\mathcal{P}}^h$
\begin{equation}
    \Pr [\mathsf{UpdateSt}(st_{e-1},pp,v_e,{\pi}_e) \neq st_e] \leq \mathsf{negl}(\lambda)
\end{equation}


\end{definition}


\begin{definition}(Public Verifiability)
Given $\mathsf{VerifyRand}(\ )$ as a public probabilistic polynomial-time algorithm run by an external verifier $P_x \notin \mathcal{P}$ that receives $v_e, {\pi}_e$ and the state $st_{e-1}$ at the end of round $e$ as input values and outputs a bit $\mathsf{0}$ or $\mathsf{1}$ based on the verification of $v_e$ using ${\pi}_e$. Then for every round $e \geq 1$
\begin{equation}
    \Pr [\mathsf{VerifyRand}(v_e,{\pi}_e,st_{e-1}) \neq \mathsf{1}] \leq \mathsf{negl}(\lambda)
\end{equation}
\end{definition}

\section{DRB Classification} \label{Sec:DRB-classification}
We classify DRB protocols in two ways: \textit{Interactive} and \textit{Non-Interactive}. \textit{Interactive} DRB protocols generate a beacon output in an interactive manner which involves multiple rounds of communication among participants. However, \textit{Non-Interactive} DRB protocols do not involve interactions among participants to produce a random beacon value for each round. Therefore, non-interactive DRBs are preferable for decentralized applications. Nevertheless, the setup for the public parameter generation can be interactive for both types of DRBs.

\subsection{Interactive Decentralized Randomness Beacon Protocols} \label{Sec:I-DRB}
Interactive DRB protocols employ multiple rounds of interaction among participants in order to produce one beacon output. These protocols are constructed using interactive cryptographic primitives such as Publicly Verifiable Secret Sharing (PVSS) or Interactive Threshold Signature Scheme. The existing interactive DRB protocols are based on PVSS involving two logical rounds of coin-tossing wherein the first round, the participants broadcast commitments to their shares, and further, these commitments are revealed in another round. Constructions of DRBs with other interactive cryptographic primitives, we left as open problems.

\begin{researchproblem}
 Construct a DRB protocol based on interactive threshold signature scheme with better complexity compared to existing interactive DRBs.
\end{researchproblem}

The main advantage of PVSS-based DRBs is that the generated randomness is indistinguishable from uniform. Nevertheless, due to the interaction and broadcast, interactive DRBs incur high communication cost. Some of the PVSS-based DRBs improve upon the general PVSS scheme to reduce the communication complexity by utilizing a threshold version of PVSS or electing a committee to perform PVSS or introducing a leader to relay the messages. Hence, these optimized versions of DRB protocols can be used to obtain periodic fresh randomness in real-world applications. A PVSS scheme consists of a tuple of algorithms $\mathsf{(PVSS.Setup, PVSS.Share, PVSS.Verify, PVSS.Recon)}$ described in Appendix~\ref{appendix:PVSS}.

PVSS-based DRB protocols are mainly of two types: with leader~\cite{schindler2020hydrand,bhat2020randpiper,das2021spurt} and without leader~\cite{kiayias2017ouroboros,cascudo2017scrape,cascudo2020albatross}. In a leader-based protocol, a leader $L_e$ is elected in each round $e$ which is responsible for performing the distribution of the secret shares of the PVSS scheme. A further illustration can be found in Appendix~\ref{appendix:PVSS}. Following we present a description of PVSS-based interactive DRB protocols. 
\begin{itemize}[leftmargin=*]
    \item Ouroboros~\cite{kiayias2017ouroboros}: Ouroboros is a PoS-based blockchain where a set of elected participants run the DRB protocol to fetch the randomness for the leader election. It operates in two phases \textit{commit} and \textit{reveal}. In \textit{commit} phase, participants encrypt the shares for all other participants by running $\mathsf{PVSS.Share}$ and submit the shares on the blockchain. In \textit{reveal} phase, each participant decrypts all the encrypted shares that are encrypted using his public key. Then, each participant computes a local random value using all the decrypted shares and posts it in the blockchain. Finally, a beacon output is computed by performing an XOR operation on all the published local random values.

    \item RandHound, RandHerd~\cite{syta2017scalable}: Syta et al. constructed scalable randomness generation protocols by following client-server architecture and threshold cryptography. RandHound is a one-shot on-demand protocol to generate single randomness. However, RandHerd is a beacon protocol that emits continuous random values. RandHound divides the servers into groups, and each group is responsible for running PVSS among the group members. RandHound employs the commit-reveal technique as defined in Ouroboros for each group. Finally, to produce global randomness in RandHound, a client operates on all the received valid local randomness from each server group. RandHerd improves upon the complexity of RandHound by leveraging communication trees among the server groups and collective signing to produce beacon outputs.

    \item SCRAPE~\cite{cascudo2017scrape}: Cascudo et al. constructed an honest majority coin-tossing protocol SCRAPE with guaranteed output delivery. It constructs a threshold PVSS scheme where sharing, verification, and reconstruction take only a linear number of exponentiations compared to quadratic in basic PVSS scheme~\cite{schoenmakers1999simple}. In SCRAPE, all participants have access to a ledger where messages are posted similar to Ouroboros. Cascudo et al. constructed an efficient share verification procedure with linear complexity by observing the fact that sharing a secret using PVSS is equivalent to encoding the secret with a Reed Solomon error correcting code~\cite{reed1960polynomial}. The dealer in the PVSS scheme~\cite{schoenmakers1999simple} not only encrypts the shares but also commits to the shares. Therefore, to prove that shares in encrypted shares are the same as shares in commitments, the efficient share verification procedure involving error-correcting code is applied. SCRAPE improves the computation and verification cost compared to Ouroboros. 
    
    
    \item HydRand~\cite{schindler2020hydrand}: HydRand improves upon the complexity of SCRAPE's PVSS protocol. HydRand works in rounds consisting of three phases: \textit{propose}, \textit{acknowledge} and \textit{vote}. In each round, a leader is selected deterministically from the set of potential leaders and by using the last round randomness. In \textit{propose} phase, the leader reveals his previously committed value which is acknowledged, signed and further broadcast by the other participants in \textit{acknowledge} phase. In \textit{vote} phase, each participant performs some checks, including the checks on the number of received acknowledgments. If the leader does not reveal his secret, the secret is reconstructed using $\mathsf{PVSS.Recon}$. The beacon value is computed using the revealed secret and the last round randomness.  
    
    
    \item ALBATROSS~\cite{cascudo2020albatross}: ALBATROSS significantly improves, amortizes the computation complexity of SCRAPE and provides a universal composability (UC)-secure model. It shows efficiency gain through the packed Shamir secret sharing scheme in PVSS or by using a linear t-resilient function to extract randomness as a vector of random group elements. It utilizes Cooley-Tukey fast Fourier transformation to amortize the complexity and for further improvement, it uses $\sum$-protocol to prove that the published sharing is correct. ALBATROSS provides two variants of UC security: 1) First variant uses UC-Non-Interactive Zero-Knowledge (NIZK) proofs for discrete logarithm, 2) Second variant introduces and uses a new primitive named ``designated verifier'' homomorphic commitments where a sender can open a commitment for one specific receiver. Later, the receiver can prove the same opening to a third party.
    
    
    \item RandPiper~\cite{bhat2020randpiper}: Bhat et al. constructed a reconfiguration-friendly DRB protocol RandPiper with strong security guarantees and quadratic communication complexity. It combines PVSS with State-Machine Replication protocol and presents two protocols: GRandPiper and BRandPiper. GRandPiper is a communication optimal DRB with strong unpredictability in the presence of a static adversary. However, BRandPiper shows the best communication complexity and the best possible unpredictability in case of a dynamic adversary. 
    
    
    \item SPURT~\cite{das2021spurt}: SPURT protocol constructs a new PVSS scheme using pairing to produce beacon output and involves a leader. The new PVSS scheme relies on Decisional Bilinear Diffie-Hellman (DBDH) assumption~\cite{boneh2001short}. In addition, SPURT uses State Machine Replication to lower the communication complexity compared to the broadcast channel used by other DRBs e.g., HydRand. SPURT operates in a semi-synchronous network and has no trusted setup.
    
\end{itemize}

\subsection{Non-Interactive Decentralized Randomness Beacon Protocols} \label{Sec:NI-DRB}
We categorize Non-Interactive DRB (NI-DRB) protocols based on the main constituent cryptographic primitive, further, we illustrate these protocols in Table~\ref{tab:NI-DRB}.  

\vspace{-0.09cm}
\subsubsection{VDF-based}
These DRBs are based on stand-alone Verifiable Delay Function $\mathsf{VDF = (VDF.Setup, VDF.Eval, VDF.Verify)}$ described in Appendix~\ref{appendix:VDF}. A VDF is a function $f:\mathcal{X} \rightarrow \mathcal{Y}$ that takes a prescribed number of sequential steps to compute the output and provides exponentially easy verification of the output. In a VDF-based DRB, the participants evaluate an Iteratively Sequential Function (ISF) to generate their local random values. The verification of these values can be efficiently done using $\mathsf{VDF.Verify}$. Due to the non-parallelizable property of VDF, an adversary cannot bias the output of the random beacon.


Lenstra and Wesolowski~\cite{lenstra2015random} constructed a DRB protocol, Unicorn, using a slow-time hash function named \textit{sloth}. This function takes inputs from a set of distrusting participants and outputs a random value. Keeping Unicorn protocol as a successor to VDF, the following VDF-based DRB protocols are constructed.


\begin{itemize}[leftmargin=*]
    \item Minimal VDF Randomness Beacon~\cite{drake2018minimal}: Justin Drake constructed a minimal randomness beacon using RANDAO~\cite{RANDAO} and VDF. RANDAO is a smart contract based DRB where participants submit their local entropy to the smart contract, and further, the smart contract produces a single global entropy. RANDAO biasable entropy is used as input to the VDF to produce unbiasable randomness. Nevertheless, there is no formal security analysis of this protocol.
    \item Continuous VDF~\cite{ephraim2020continuous}: Ephraim et al. presented a new notion of Continuous Verifiable Delay Function (cVDF) by adapting Pietrzak scheme~\cite{pietrzak:LIPIcs:2018}. A cVDF $f$ provides the output computation of each intermediate steps (i.e. $f(t)$ for $t < T$) with an efficient proof $\pi^t$ (used for public verification of the output). A cVDF can be used to construct a DRB protocol where beacon outputs are generated by applying a suitable hash to the intermediate outputs of each step. The drawback with this protocol is that the nodes having the most efficient (fastest) processors can always learn the beacon outputs before the other participating nodes. A similar argument goes for the Unicorn protocol.
    \item RandRunner~\cite{schindler2021randrunner}: RandRunner leverages trapdoor VDF with strong uniqueness to construct a DRB protocol. Each participant $P_i$ of RandRunner initializes its public parameter ${pp}_i$ with a corresponding trapdoor $sk_i$. The participants exchange their public parameters and verify the received ones. RandRunner executes in consecutive rounds where in each round, a leader is elected. Further, the leader tries to solve the VDF using its trapdoor, and other participants attempt to solve the VDF using the common $\mathsf{VDF.Eval}$ algorithm. The drawback with the RandRunner protocol is that once a powerful adversary becomes a leader, it can keep corrupting the round leaders (e.g., via DoS), withhold its output computed via trapdoor, and keep working on for the next outputs for many subsequent rounds hence breaking unpredictability. 
    
    \item RANDCHAIN~\cite{han2020randchain}: RANDCHAIN is a competitive DRB where in each round, nodes compete to be a leader which solely produces the beacon output. RANDCHAIN constructs a non-parallelizable Sequential Proof-of-Work (SeqPoW) puzzle by employing VDF or Sloth. A node solves the SeqPoW puzzle by incrementing an ISF for a randomized time. RANDCHAIN works as a Nakamoto-based blockchain where nodes synchronize their local blockchains and keep solving the puzzle to mine new blocks to the main blockchain. RANDCHAIN mimics a blockchain structure, so it can suffer from front-running (block withholding) attacks and can also have forks due to problems with blocks' finality.
\end{itemize}


\subsubsection{VRF-based}
These DRBs compute randomness using Verifiable Random Function $\mathsf{VRF = (VRF.KeyGen, VRF.Eval, VRF.Verify)}$ described in Appendix~\ref{appendix:VRF}. A VRF is a pseudorandom function that produces pseudorandom output along with proof about the correctness of the output. Participants in these DRBs apply VRF on an input seed to generate their local entropy which is used to compute the beacon output. VRF-based DRBs are explained as follows:



\begin{itemize}[leftmargin=*]
    \item Blockchain Protocol Designs: Ouroboros Praos~\cite{david2018ouroboros}, Algorand~\cite{gilad2017algorand} and Erlond~\cite{elrond2019highly} blockchains have their DRB as a byproduct. In these DRBs, each participant $P_i$ runs $\mathsf{VDF.Eval}$ on a seed (e.g., previous output or state) using its secret key $sk_i$ and the DRB output is computed from the participants' VRF outputs. These DRBs do not guarantee generation of uniformly random values and do not have strong bias-resistance as an adversary can include/exclude the corrupted participants' VRF outputs used for DRB output computation.
    

    \item Distributed VRF-based DRBs: A distributed VRF (DVRF)~\cite{dodis2003efficient} based DRB was first introduced by DFINITY~\cite{hanke2018dfinity}. Later DRBs~\cite{Corestar,DAOBet} employed DFINITY-DVRF along with BLS cryptography. Nevertheless, these DRBs do not provide formal security analysis. A recent paper~\cite{galindo2020fully} provides two new constructions of DVRF: 1) DDH-DVRF based on elliptic curve cryptography; 2) GLOW-DVRF based on cryptographic pairings. These constructions also formalize a security model with proper security analysis. DRBs based on DDH-DVRF, and GLOW-DVRF show strong bias resistance and strong pseudorandomness.

    \item RandChain~\cite{wang2020randchain}: RandChain follows \textit{commit-and-reveal} strategy by building a sub-routine \textit{RandGene} using VRF. RandChain has a two-layer hierarchical blockchain structure where nodes form distinct committees. Each committee has a local blockchain and generates local entropy through the RandGene protocol, further, global randomness is computed from these local entropy by forming a RandChain block. RandChain security depends on a secure sharding process, followed by a leader election for each shard (committee). However, both processes can be influenced by an adversary to obstruct DRB properties.
    
    
\end{itemize}



\vspace{-0.6cm}
\subsubsection{HE-based}
These DRBs utilize homomorphic encryption scheme $\mathsf{HE} = (\mathsf{HE.Setup}$,\\ $\mathsf{HE.KeyGen, HE.Enc, HE.Dec, HE.Eval)}$. Homomorphic encryption allows performing arithmetic operations on ciphertext directly without decryption (details in Appendix~\ref{appendix:HE}). Following DRBs employ ElGamal encryption~\cite{elgamal1985public} as partial HE.


\begin{itemize}[leftmargin=*]
    \item Nguyen-Van et al.~\cite{nguyen2019scalable}: Their DRB has three components: a Requester, a Public Distributed Ledger (PDL), and a Core Layer. The protocol works in rounds where, first, the Requester sends a nonce to the PDL that computes a ticket $T$ and publishes it. Further, participants of the core layer run a VRF using the ticket $T$ to check if they are selected as a contributor. Each contributor publishes a ciphertext computed on a random value using the Requester's public key. Later, the Requester performs a homomorphic operation on the published ciphertexts and computes a single ciphertext. Finally, the Requester publishes the decrypted value as DRB output with a proof of correct decryption. There are two drawbacks: 1) A malleable ElGamal encryption, 2) The Requester can collude with contributors or refuse to decrypt the resulting ciphertext. 
    
    \item HERB~\cite{cherniaeva2019homomorphic}: Homomorphic Encryption Random Beacon (HERB) DRB uses threshold ElGamal encryption scheme with a distributed key generation (DKG) protocol. DKG is used to generate a common public key and secret key shares for participants. Each participant publishes a ciphertext share with proof of correct encryption (NIZK Proof) on a public bulletin board. These shares generate an aggregated ciphertext through ElGamal aggregation which is subsequently decrypted by a threshold of participants to produce the DRB output.

\end{itemize}



\vspace{-0.6cm}
\subsubsection{External Source-based}
In these DRBs, participants extract the randomness from an external entropy source, i.e., real-world entropy. These entropy sources can be public blockchains~\cite{bunz2017proofs,bentov2016bitcoin}, real-time financial data~\cite{clark2010use} or national lottery results~\cite{baigneres2015trap}. PoW-based blockchains are promising sources but an adversarial miner can manipulate the generated randomness. Therefore, to achieve most of the beacon properties, the following DRBs apply different defense mechanisms.

\begin{table*}[bth]
    \resizebox{\textwidth}{!}{
    \centering
    \begin{tabular}{|l|c|c|}
        \hline
        \textbf{Scheme} & \textbf{Advantages}  & \textbf{Disadvantages}\\
        \hline

        VDF-based & \makecell[l]{1. These DRBs achieve liveness under the period of full asynchrony. \\2. These DRBs avoid byzantine agreement consensus hence have \\less communication complexity. \\3.  It shows strong bias-resistance as long as there is an honest node.} & \makecell[l]{1. Front-running attack can hinder some DRB properties.\\ 2. In most of these DRBs, the significant powerful adversary\\ can learn the output of DRB earlier than other nodes. \\ 3. These protocols rely on the new assumptions of VDF.} \\[0.1cm]
        \hline 
        
        VRF-based & \makecell[l]{1. Most of these DRBs do not have any trusted setup and achieve \\strong notion of pseudo-randomness and bias-resistance~\cite{galindo2020fully}. \\ 2. These DRBs incur less computation and communication cost.} & \makecell[l]{1. In some of these DRBs, leader uniqueness is not \\guaranteed that introduces additional consensus protocol \\to agree on the beacon output.}\\[0.1cm]
        \hline 
        
        HE-based & \makecell[l]{1. The output of these DRBs for a round $e$ does not depend on the\\ output of the previous round $e-1$. \\ 2. Partial homomorphic encryption schemes used in these DRBs \\can be replaced by a lattice-based fully homomorphic scheme to\\ ensure the post-quantum security.} & \makecell[l]{1. Scalability issue due to the homomorphic evaluation of \\multiple ciphertexts. \\ 2. The existing DRBs use public ledger to publish the local\\ and global entropy. But distributing the local entropy in\\ DRBs using a consensus incur a high communication cost.}\\[0.1cm]
        \hline 
        
        \makecell[l]{External\\Source\\-based} & \makecell[l]{1. These DRBs do not incur communication cost as the DRB \\output is published in a public bulletin board.\\ 2. These DRBs work perfectly even in the asynchronous network.} & \makecell[l]{1. Most of these DRBs do not provide public verifiability. \\ 2. Proof-of-Work based beacons are not energy efficient and\\ nodes with better hardware can outperform other nodes in\\ producing the beacon output.}\\[0.1cm]
        \hline 
        
        \makecell[l]{Threshold\\Signature\\-based} & \makecell[l]{1. These DRBs provide strong bias resistance and unpredictability. \\ 2. Consortium of organizations can participate to construct\\ such beacon due to threshold property (e.g. Drand~\cite{Drand}).} & \makecell[l]{1. These DRBs require either a trusted setup or DKG, hence \\  do not offer a reconfiguration-friendly setup. \\ 2. Security of the DRBs depend on the security assumptions \\ of elliptic curve pairings due to the use of BLS-signature.}\\[0.1cm]
        \hline
        
    \hline        
    \end{tabular}}
    \caption{Advantages and Disadvantages of different Non-Interactive DRB protocols}
    \label{tab:NI-DRB}
    \vspace{-0.82cm}
    \end{table*}
 
\begin{itemize}[leftmargin=*]
    \item Rand Extractor~\cite{clark2010use}: Clark et al.~\cite{clark2010use} created a model to generate randomness by combining the information theory with computational finance. They used the closing prices of the stock market to compute a random output. During the market's closing in the day, one entity publishes this random output in the protocol. This entity can also induce its own local entropy to transparently construct a publicly verifiable final randomness, but liveness is hard to achieve.
    
    
    \item Proofs of Delay~\cite{bunz2017proofs}: In this DRB, a beacon smart contract (BC) publishes the random beacon values on a public blockchain. The DRB is built on \textit{Proof-of-Delay} which uses an ISF such as sloth~\cite{lenstra2015random}. In this DRB, a beacon maintainer executes this ISF and publishes the result to BC with queryable access to the beacon output using a \textit{refereed delegation of computation} protocol. To show the honest behavior, the maintainer is incentivized; otherwise punished.

    
    \item Bitcoin Beacon~\cite{bonneau2015bitcoin}~\cite{bentov2016bitcoin}: These DRBs extract randomness from the bitcoin blockchain~\cite{Nakamoto_bitcoin} and follows the security of the bitcoin. In~\cite{bonneau2015bitcoin}, an extractor fetches the randomness from the block headers. As each block contains several transactions involving ECDSA signatures~\cite{gennaro2016threshold} that rely on strong randomness for security hence, the extractor gives good public randomness as a beacon output. Bentov et al.~\cite{bentov2016bitcoin} constructed a bitcoin beacon protocol that fetches $m$ consecutive blocks $B_1,B_2,\ldots,B_m$ such that the block $B_m$ already have $l$ subsequent blocks. Further, the protocol acquires a bit $b_i$ from each block and runs a $\mathsf{majority}$ function on all these bits as input to get the DRB output. 
    
\end{itemize}


\vspace{-0.6cm}
\subsubsection{Threshold Signature-based}
These DRBs are based on a non-interactive threshold signature scheme that requires a single round of communication among participants to produce the unique group signatures from a threshold number of participants' signature shares. Most of the existing threshold signature-based DRBs employ threshold BLS signature. These DRBs require a setup to generate the secret shares for the participants. Additionally, the complexity of unique signature construction comply with DRB protocol for practical use.

\begin{itemize}[leftmargin=*]
    \item Cachin et al.~\cite{cachin2005random}, Drand~\cite{Drand}: Cachin et al. presented a common coin protocol using threshold signature along with a random-access coin-tossing scheme. In this DRB, a trusted dealer distributes the secret key shares to the participants. The DRB output is a unique signature on the hash of a counter (epoch number). Drand~\cite{Drand} follows a similar idea, but it replaces the threshold secret to the threshold BLS key. Drand can be considered as an implementation of the Cachin et al. scheme. Drand utilizes the DKG protocol of Gennaro et al.~\cite{gennaro1999secure} during the setup phase that yields a high communication complexity.
    
    \item DFINITY~\cite{hanke2018dfinity}: It also employs a threshold BLS signatures scheme but the selection of the best initialization vector in the scheme creates a challenge. The protocol works well even in the partial synchronous network model. It employs a non-interactive DKG setup and achieves better communication complexity than Drand. The DRB acts as a VRF that produces unbiasable output.
    
\end{itemize}
\textbf{Note:} We present Hybrid DRB protocols in Appendix~\ref{appendix:other-DRBs}.


\section{Discussion} \label{Sec:Discussion}

\subsection{Security Assumptions}
The security of all these DRBs depends on well-defined security assumptions. These assumptions can be assumptions about the underlying network, adversary, setup, or cryptographic primitives. If these assumptions are failed in some cases, then the DRB using these assumptions will break its security properties.
\begin{itemize}[leftmargin=*]
    \item \textit{Cryptographic Assumptions (Primitive)}: As the above described DRBs are based on cryptographic primitives such as PVSS, VDF, VRF, these DRBs inherit the security assumptions from the primitives. These assumptions are well-known hard problems of cryptography such as standard decisional or computational Diffie-Hellman assumptions~\cite{boneh1998decision} (or their variants) depending on the underlying cryptographic scheme (e.g., PVSS, DVRF). VDF-based DRBs depend on the new security assumptions on sequential computation (e.g., iterated squaring over groups of unknown order~\cite{rivest1996time}) that are not well studied and understood in the current literature. Modeling of the hash function as a random oracle~\cite{bellare1993random} is also considered in security assumptions in some DRBs. 
    \item \textit{Network Assumptions (Model)}: Most of the PVSS-based and VRF-based DRBs assume a strong \textit{synchronous} network which can be an unrealistic setting in the real world. Hence, these DRBs require a lock-step synchronous network where the messages are delivered before the end of each round. In case of no lock-step synchrony, participants might employ round synchronization protocols~\cite{naor2019cogsworth,yin2019hotstuff}. Some of the DRBs work well in \textit{semi-synchronous} network where the messages are delivered within a known finite time-bound. VDF-based and external-source-based DRBs work well in an \textit{asynchronous} network where messages are delivered without a known time-bound. However, the trust of these models depends on the underlying setup assumptions or on the public blockchain, where the local entropy of the participants are posted.
    \item \textit{Setup Assumptions}: Many DRB protocols~\cite{Drand,cherniaeva2019homomorphic,bhat2020randpiper,syta2017scalable,hanke2018dfinity} require an initial trusted setup assumption where private keys for the participants and uniformly random public parameters are generated by a trusted third party (dealer) or by a distributed key generation (DKG) protocol. The Security of DRBs with a trusted third party crucially depends on the action and ability of the trusted party. Nevertheless, DKG incurs a considerable setup cost (high communication complexity) with its limitation of adding or replacing the participants. Therefore, DKG-based DRBs are preferred when the participants are fixed. Hence, many recent DRB protocols~\cite{schindler2020hydrand,cascudo2017scrape,david2018ouroboros,azouvi2018winning,gilad2017algorand} have a transparent setup where the public parameters are trapdoor free. 
\end{itemize}

Following the above security assumption, most of the DRB protocols perform well in permissioned systems. However, permissionless systems have a highly dynamic set of nodes that maintain the system state. Due to the dynamically changing participants, integrating an existing DRB with the system is challenging. Moreover, setting the assumption on a number of adversarial nodes is hard.

\begin{researchproblem}
 Study the hardness of embedding the existing DRB protocols in permissionless systems, based on Proof-of -Work (PoW) or -Stake (PoS).
\end{researchproblem}

\subsection{Complexity}
DRB protocols following different approaches exhibit different complexity. Finding a good balance between computation and communication complexity in a DRB protocol is a challenging task. Therefore, an extensive amount of work has been devoted to reduce the complexity of DRB protocols. 
\begin{itemize}[leftmargin=*]
    \item \textit{Communication Complexity}: Most of the interactive DRB protocols assume a broadcast channel. Therefore, Ouroboros~\cite{kiayias2017ouroboros}, RandShare~\cite{syta2017scalable}, and SCRAPE \cite{cascudo2017scrape} have a communication complexity of $\mathcal{O}(n^3)$ due to the broadcasting of  $\mathcal{O}(n)$ size message. HydRand~\cite{schindler2020hydrand} improves upon the communication complexity to $\mathcal{O}(n^2)$ by having a leader-based approach where a leader node performs the PVSS share distribution. Relaying the messages through a single node to reduce the communication complexity is also embraced by ALBATROSS~\cite{cascudo2020albatross}, GLOW~\cite{galindo2020fully}. RandHound, RandHerd~\cite{syta2017scalable}, DFINITY~\cite{hanke2018dfinity} employ sharding to sample a committee for output generation that results in lower communication complexity. But such a procedure can be immediately subject to attacks by an adaptive adversary who can corrupt the committee once it is determined.
   
    Most of the non-interactive DRBs~\cite{david2018ouroboros,gilad2017algorand,bunz2017proofs,nguyen2019scalable} have less communication complexity as a successful participant (e.g., leader) usually need to perform one broadcast. Therefore, it incurs the communication complexity in $\mathcal{O}(n)$. Moreover, most of the NI-DRBs involving blockchain~\cite{RANDAO,han2020randchain} to publish shared local and global randomness also have lower communication complexity. 
    
    DKG setup based DRBs~\cite{cherniaeva2019homomorphic,hanke2018dfinity,Drand,cachin2005random} suffer from additional communication cost. Complexity of DRBs can be improved using asynchronous data dissemination (ADD)~\cite{dasasynchronous} or using hbACSS~\cite{yurek2021hbacss}(for PVSS-based DRBs).
    \begin{researchproblem}
     Design a DRB protocol with $\textit{sub-quadratic}$ communication complexity together with optimal fault-tolerance.
    \end{researchproblem}
    \item \textit{Computation Complexity}: It is defined as the number of operations needed to be performed by a participant during one round of DRB protocol. PVSS-based protocols such as RandShare~\cite{syta2017scalable} and Ouroboros~\cite{kiayias2017ouroboros} requires a computation complexity of $\mathcal{O}(n^3)$. An improved version of PVSS further reduces this cost in SCRAPE~\cite{cascudo2017scrape}. Puzzle-based DRBs ~\cite{bonneau2015bitcoin,bunz2017proofs} have a high computational cost due to the involved puzzle. VDF-based DRBs also have the drawback of high computational complexity due to the repeated squaring. On the contrary, VRF-based DRBs incur a minimum computation cost. 
    \begin{researchproblem}
     Design a puzzle-based DRB protocol incurring low computation complexity.
    \end{researchproblem}
    
    \item \textit{Verification Complexity} Verification cost refers to the number of operations performed by an external participant to verify the output of a beacon protocol. Although VDF-based DRBs have high computational costs, they do provide efficient verification hence incur less verification cost. The most efficient DRB protocols with regard to computation and verification complexity are based on VRF~\cite{david2018ouroboros,gilad2017algorand,wang2020randchain} or threshold crypto-systems~\cite{cachin2005random,hanke2018dfinity,Drand,nguyen2019scalable}. 
    \begin{researchproblem}
     Design a PVSS-based DRB protocol with a constant verification complexity, linear communication cost and no trusted setup.
    \end{researchproblem}
    
\end{itemize}

\subsection{Scalability}
Despite a decade of research on DRB protocols, only quite a few recent DRBs emphasize the scalability of their DRB. Scalability in a DRB protocol refers to the number of participants it can support. Many of the described DRB protocols do not offer good scalability. Especially, DRBs involving DKG setup provide poor scalability as DKG does not support frequent modification in the set of key holders. In addition, the high complexity along with the underlying network model in many of these DRBs significantly affect the scalability of the DRBs. 

A general approach for achieving good scalability is ``\textit{sharding}'' which is considered in recent DRBs, including RandHerd~\cite{syta2017scalable}, DFINITY~\cite{hanke2018dfinity} and Algorand~\cite{gilad2017algorand} with the cost of slightly degrading the fault-tolerance. RandHerd shows a direct consequence of the sharding where nodes are split into smaller groups. Each group produces local entropy and each group's entropy is combined to produce the DRB output. Algorand and DFINITY show selection of a committee to generate the DRB output. Therefore, this orthogonal technique of randomly sampling a committee for protocol execution can improve scalability.


Another way for improving scalability is using a \textit{leader-based approach} where a leader relays the messages to the participants. Moreover, having a \textit{public ledger} where participants post their local entropy messages also improves scalability.

\textit{Reconfiguration Friendliness} directly impact scalability. A protocol is reconfiguration friendly when the parameters and list of participants can be changed dynamically without affecting the current execution. When there is no binding between the setup and the system, the reconfiguration becomes easier. DRBs involving DKG setup are not reconfiguration-friendly, hence poor scalability. On the contrary, non-interactive DRBs (with no DKG) show better scalability.


\begin{researchproblem}
 Study the (im)possibility of designing a reconfiguration-friendly sub-quadratic DRB protocol that do not employ committee sampling.
\end{researchproblem}

 \begin{table*}[htbp]
   
    \resizebox{\textwidth}{!}{
    \centering
    \begin{threeparttable}
     \caption{Comparison of Existing Decentralized Randomness Beacon Protocol}
    \label{tab:comparison}
    \begin{tabular}{l c c c c c c c c c c c}
    \hline
    & & & & & & & & & & &\\
    \rotatebox{90}{Protocol} & \rotatebox{90}{Network Model} & \rotatebox{90}{Adaptive Adversary} & \rotatebox{90}{Liveness} & \rotatebox{90}{Unpredictability} & \rotatebox{90}{Bias-Resistance} & \rotatebox{90}{Fault-tolerance} & \rotatebox{90}{\makecell{Communication \\ Complexity}} & \rotatebox{90}{\makecell{Computation \\ Complexity}} & \rotatebox{90}{\makecell{Verification \\ Complexity}} & \rotatebox{90}{\makecell{Cryptographic \\ Primitive}} & \rotatebox{90}{\makecell{No Trusted Dealer \\ or DKG required}}
     \\[0.1cm]
    \hline
     ALBATROSS~\cite{cascudo2020albatross} & syn. & \xmark & \cmark & \cmark & \cmark & 1/2 & $\mathcal{O}(n)$ & $\mathcal{O}(\log n)$ & $\mathcal{O}(n)$ & PVSS & \cmark \\[0.1cm]
    
     Algorand~\cite{gilad2017algorand} & semi-syn. & \xmark & \cmark & \cmark$^{\ddagger}$ & \xmark & $1/3^{\diamond}$ & $\mathcal{O}(cn)$ & $\mathcal{O}(c)$ & $\mathcal{O}(1)$ & VRF & \cmark \\[0.1cm] 
    
     BRandPiper~\cite{bhat2020randpiper} & syn. & \cmark & \cmark & \cmark & \cmark & 1/2 & $\mathcal{O}(n^3)$ & $\mathcal{O}(n^2)$ & $\mathcal{O}(n^2)$ & PVSS & \cmark \\[0.1cm]

    Cachin et. al~\cite{cachin2005random} & asyn. & \xmark & \cmark & \cmark & \cmark & 1/3 & $\mathcal{O}(n^2)$ & $\mathcal{O}(n)$ & $\mathcal{O}(1)$ & Uniq. thr-sig. & \xmark \\[0.1cm]
    
    Caucus~\cite{azouvi2018winning} & syn. & \xmark & \cmark & \cmark$^{\ddagger}$ & \xmark & 1/2 & $\mathcal{O}(n)$ & $\mathcal{O}(1)$ & $\mathcal{O}(1)$ & Hash func. & \cmark \\[0.1cm]
    
    Continuous VDF~\cite{ephraim2020continuous} & asyn. & \xmark & \xmark$^{\dagger}$ & \cmark & \cmark & 1/2 &  $\mathcal{O}(1)$ & VDF &  $\mathcal{O}(1)$ & VDF & \cmark \\[0.1cm]

    DFINITY~\cite{hanke2018dfinity} & semi-syn. & \xmark & \cmark & \cmark & \cmark & 1/3 & $\mathcal{O}(n^2)$ & $\mathcal{O}(n)$ & $\mathcal{O}(1)$ & BLS thr-sig. & \xmark \\[0.1cm]
    
    Drand~\cite{Drand} & syn. & \xmark & \cmark & \cmark & \cmark & 1/2 & $\mathcal{O}(n^2)$ & $\mathcal{O}(n)$ & $\mathcal{O}(1)$ & Uniq. thr-sig. & \xmark \\[0.1cm]
    
    GLOW~\cite{galindo2020fully} & syn. & \xmark & \cmark & \cmark & \cmark & 1/3 & $\mathcal{O}(n)$ & $\mathcal{O}(n)$ & $\mathcal{O}(1)$ & DVRF & \xmark  \\[0.1cm]
    
    GRandPiper~\cite{bhat2020randpiper} & syn. & \xmark & \cmark & \cmark$^{\ddagger}$ & \cmark & 1/2 & $\mathcal{O}(n^2)$ & $\mathcal{O}(n^2)$ & $\mathcal{O}(n^2)$ & PVSS & \cmark \\[0.1cm]
    
    HERB~\cite{cherniaeva2019homomorphic} & syn. & \xmark & \cmark & \cmark & \cmark & 1/3 & $\mathcal{O}(n^2)^{\ast}$ & $\mathcal{O}(n)$ & $\mathcal{O}(n)$ & PHE & \xmark \\[0.1cm]
    
    HydRand~\cite{schindler2020hydrand} & syn. & \xmark & \cmark & \cmark$^{\ddagger}$ & \cmark & 1/3 & $\mathcal{O}(n^2)$ & $\mathcal{O}(n)$ & $\mathcal{O}(n)$ & PVSS & \cmark \\[0.1cm]
    
     Nguyen-Van et. al~\cite{nguyen2019scalable} & syn & \xmark & \xmark & \cmark & \xmark & 1/2 & $\mathcal{O}(n)$ & $\mathcal{O}(1)$ & $\mathcal{O}(n)$ & PHE, VRF & \cmark \\[0.1cm]
    
    Ouroboros~\cite{kiayias2017ouroboros} & syn. & \xmark & \cmark & \cmark & \cmark & 1/2 & $\mathcal{O}(n^3)$ & $\mathcal{O}(n^3)$ & $\mathcal{O}(n^3)$ & PVSS & \cmark \\[0.1cm]
    
    Ouroboros Praos~\cite{david2018ouroboros} & semi-syn. & \cmark & \cmark & \cmark$^{\ddagger}$ & \xmark & 1/2 &  $\mathcal{O}(n)^{\star}$ & $\mathcal{O}(1)^{\star}$ & $\mathcal{O}(1)^{\star}$ & VRF & \cmark \\[0.1cm]
    
    Proof-of-Delay~\cite{bunz2017proofs} & syn. & \xmark & \cmark & \cmark & \cmark & 1/2 & $\mathcal{O}(n)$ & very high & $\mathcal{O}(\log{\Delta})^{\circ}$ & Hash func. & \cmark \\[0.1cm]
    
    Proof-of-Work~\cite{Nakamoto_bitcoin} & syn. & \xmark & \cmark & \cmark$^{\ddagger}$ & \xmark & 1/2 & $\mathcal{O}(n)$ & very high & $\mathcal{O}(1)$ & Hash func. & \cmark  \\[0.1cm]
    
    RandChain~\cite{wang2020randchain} & syn. & \xmark & \cmark & \cmark & \cmark & 1/3 & $\mathcal{O}(cn)$ & $\mathcal{O}(cn)$ & $\mathcal{O}(n)$ & VRF & \cmark \\[0.1cm]
    
    RANDCHAIN~\cite{han2020randchain} & syn. & \cmark & \cmark & \cmark & \cmark & 1/3 & $\mathcal{O}(n)$ & VDF & $\mathcal{O}(1)$ & VDF & \cmark \\[0.1cm]
    
    RANDAO~\cite{RANDAO} & asyn. & \xmark & \cmark & \xmark & \xmark & 1/2 & $\mathcal{O}(n)$ & VDF & $\mathcal{O}(1)$ & VDF & \cmark \\[0.1cm]
    
    RandHerd~\cite{syta2017scalable} & syn. & \xmark & \cmark & \cmark & \cmark & 1/3 & $\mathcal{O}(c^{2}\log n)$ & $\mathcal{O}(c^{2}\log n)$ & $\mathcal{O}(1)$ & PVSS, CoSi & \xmark \\[0.1cm]
    
    RandHound~\cite{syta2017scalable} & syn. & \xmark & \cmark & \cmark & \cmark & 1/3 & $\mathcal{O}(c^{2}n)$ & $\mathcal{O}(c^{2}n)$ & $\mathcal{O}(c^{2}n)$ & PVSS & \cmark \\[0.1cm]
    
    RandRunner~\cite{schindler2021randrunner} & syn. & \cmark & \cmark & \cmark$^{\ddagger}$ & \cmark & 1/2 & $\mathcal{O}(n^2)$ & VDF &  $\mathcal{O}(1)$ & VDF & \cmark \\[0.1cm]
    
    RandShare~\cite{syta2017scalable} & asyn. & \cmark & \xmark$^{\odot}$ & \cmark & \cmark & 1/3 & $\mathcal{O}(n^3)$ & $\mathcal{O}(n^3)$ & $\mathcal{O}(n^3)$ & VSS & \cmark \\[0.1cm]
    
    Rand Extractor~\cite{clark2010use,bonneau2015bitcoin} & asyn.$^{\pm}$ & \cmark &  \cmark$^{\amalg}$ & \cmark & \cmark & 1/2 &  $\mathcal{O}(1)$ &  $\mathcal{O}(1)$ &  $\mathcal{O}(1)$ & Hash func. & \cmark \\[0.1cm]
    
    SCRAPE~\cite{cascudo2017scrape} & syn. & \xmark & \cmark &  \cmark & \cmark & 1/2 & $\mathcal{O}(n^3)$ & $\mathcal{O}(n^2)$ & $\mathcal{O}(n^2)$ & PVSS & \cmark \\[0.1cm]
    
    SPURT~\cite{das2021spurt} & semi-syn & \xmark & \cmark &  \cmark & \cmark & 1/3 & $\mathcal{O}(n^2)$ & $\mathcal{O}(n)$ & $\mathcal{O}(n)$ & PVSS, Pairing & \cmark \\[0.1cm]
    
    Unicorn~\cite{lenstra2015random} & asyn. & \xmark & \cmark$^{\dagger}$ & \cmark & \cmark  & 1/2 & $\mathcal{O}(1)$ & high & $\mathcal{O}(1)$ & Sloth & \cmark \\[0.1cm]
    
    \hline
    \end{tabular}
    \small{
    \begin{tablenotes}
    \item Fault-tolerance refers to number of byzantine faults a DRB can tolerate and $c$ is average committee size.
    
    \item $\ddagger$ refers to probabilistic guarantees for unpredictability, and has a bound on the number of future rounds an adaptive rushing adversary can predict the beacon output.
    
    \item $\diamond$ Due to the randomly sampling a committee of size c in Algorand, the fault-tolerance reduces slightly.
    
    \item $\dagger$ The node with more computational power learns the beacon output earlier than others.
    
    \item $\ast$ HERB achieves communication complexity of $\mathcal{O}(n^2)$ when nodes use Avalanche algorithm or public blockchain to share their ciphertexts.
    
    \item $\star$ Ouroboros Praos is not a stand-alone DRB and does not describe randomness generation approach, so the presented complexity does not account the additional complexity for communication or verification. 
    
    \item $\circ$ The verification in beacon smart contract has complexity $\mathcal{O}(\log{\Delta})$ in the security parameter $\Delta$.
    
    \item ${\odot}$ An additional synchrony assumption is needed to provide liveness in RandShare.
    
    \item $\pm$ DRB protocols built on public blockchain also follow the network structure of the respective blockchain. Therefore, ~\cite{bonneau2015bitcoin} uses a synchronous network as it relies on the bitcoin blockchain.
    
    \item $\amalg$ Liveness can be hindered due to the limited availability of financial data caused by closed exchanges in~\cite{clark2010use} or due to the fork situation in the blockchain in~\cite{bonneau2015bitcoin}.

    \end{tablenotes}}
    \end{threeparttable}}
    
\end{table*}

\vspace{-0.4cm}
\subsection{Adversarial Model}
Most of these DRBs consider a fixed set of $n$ nodes; out of these nodes, $f$ nodes may exhibit byzantine behavior. An adversary in these DRBs can be defined as:

\textit{Active vs. Passive} An active adversary actively modifies the messages (e.g., public shares, DRB output) in DRB; a passive adversary observes the transcript (i.e. messages) of an honest run of the DRB  and predicts the DRB's next output.

\textit{Adaptive vs. Static} An adaptive adversary corrupts the nodes during the protocol execution, while a static adversary does corruption before the execution. 

An adversary can affect the security guarantees of a DRB in many ways, such as by trying to bias the produced random output, withholding the output, predicting the future output, or tricking an outsider (third party) into accepting invalid beacon output. Leader-based DRBs suffer from targeted attacks, however, blockchain-based DRBs suffer from blockchain-specific attacks. Moreover, unpredictability can be affected by network model. 

\begin{researchproblem}
 Choose a static secure DRB protocol and transform it to an adaptively secure DRB that retains the efficiency standard of the static one.
\end{researchproblem}


\begin{figure}[th]
    \centering
    \includegraphics[width = 0.8\textwidth, height = 0.35\textwidth]{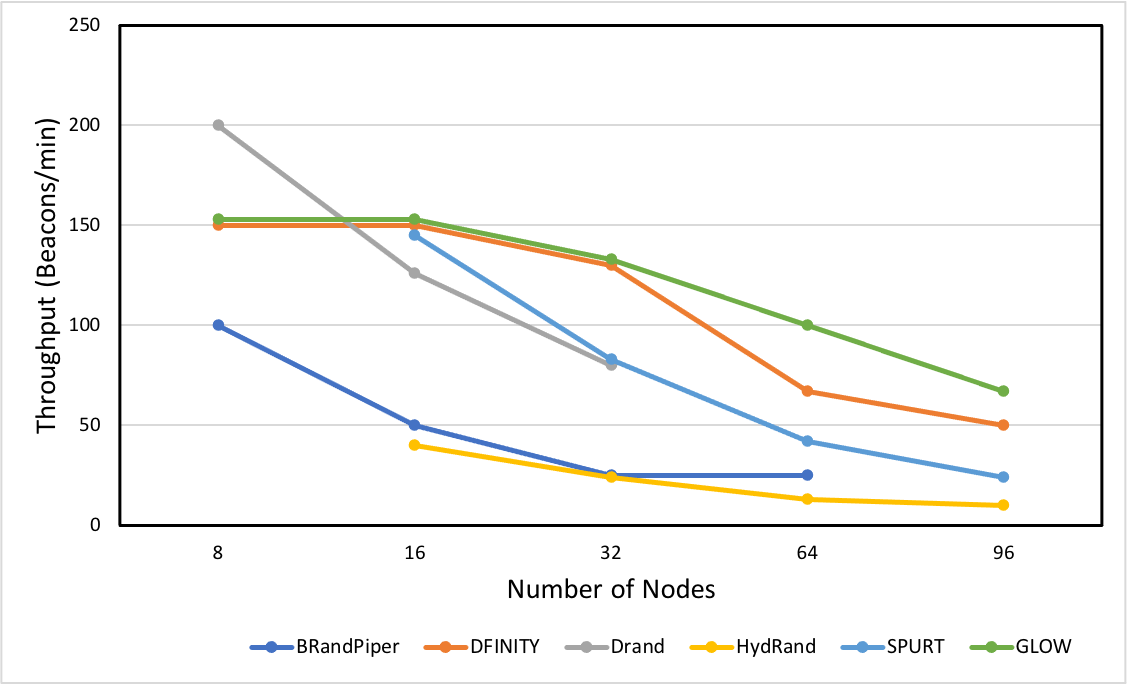}
    \vspace{-0.4cm}
    \caption{Overview of throughput for various DRB protocols}
    \label{fig:DRB}
    \vspace{-0.4cm}
\end{figure}
\subsection{Throughput Evaluation} We report throughput of state-of-the-art public implementations of various DRBs in Figure~\ref{fig:DRB}. DFINITY and SPURT operate on a semi-synchronous network but BRandPiper, Drand, GLOW, and HydRand assume synchronous networks. The network delay parameter in these DRBs directly affects the throughput of DRBs.

Drand is a practically deployed DRB protocol. However, when the number of nodes increases to more than 64, nodes in the DRand abort the DKG step of the protocol, and yet it suffers from significant network delay. For HydRand, we chose the public implementation~\cite{HydRand-code} available on Github. For BRandPiper, we depict throughput for the Merkle-tree-based implementation, which is quantitatively practical for real-world scenarios. For SPURT, we used the throughput values directly from their paper. For DFINITY and GLOW, we followed the public implementation~\cite{DVRF-code} of DVRFs to get the throughput while assuming no failure.


\vspace{-0.3cm}
\subsection{Others}
    \begin{itemize}[leftmargin=*]
    \item \textit{Incentive} Some DRBs~\cite{bunz2017proofs,RANDAO} involve incentivizing or punishing the participants to enforce fairness against rational adversaries. In particular, these (dis)incentivizing approaches (e.g.,~\cite{baum2020insured}) are considered mostly in smart contract-based DRBs. Putting an economic incentivization scheme to reward the participants of beacon enforces the honest behavior from the participants. An incentivization scheme can also reward the right computation or verification. Interesting research would be to create an \textit{incentive structure} for a DRB.
    
    \item \textit{Output Uniqueness} It states that the DRB produces a unique output even in the presence of an adversary having the trapdoor information of honest participants. It implies strong bias-resistance in DRBs. Therefore, DRBs such as RandRunner~\cite{schindler2021randrunner}, DFINITY~\cite{hanke2018dfinity} and GLOW~\cite{galindo2020fully} provide strong bias-resistance due to their output uniqueness. Having this property also prevents an adversary from manipulating the beacon output for any financial gain. 
    
    \item \textit{Universal Composability} (UC) It is arguably one of the strongest security guarantees. A UC-secure protocol ensures that the protocol can be employed as a building block in more complex systems while preserving its security. The earlier UC-secure DRBs~\cite{kiayias2017ouroboros} do not provide bias-resistance. The first UC-secure DRB ALBATROSS~\cite{cascudo2020albatross} leverages UC-secure NIZK proofs. UC-secure time-lock puzzles (TLP)~\cite{baum2021tardis} can be scrutinized to construct a UC-secure DRB.
    
\end{itemize}


\section{New Components for Construction of DRB Protocols}
There have been many new efficient constructions of cryptographic primitives in recent years. These primitives can be embedded as new building blocks or replace old ones in the DRB protocols to improve the performance of DRBs.
\begin{itemize}[leftmargin=*]
    \item \textit{Using New Verifiable Functions}: 
    Gurkan et al.~\cite{gurkan2021aggregatable} constructed a new aggregatable DKG scheme that leverages gossip instead of broadcast communication to reduce the communication complexity. Further, they introduced an efficient Verifiable Unpredictable Function (VUF) and combined it with DKG. This threshold VUF can be utilized to construct a DRB protocol. 
    
    VDF-based DRBs can benefit from the recent work~\cite{rotem2021simple} about batch verification of VDF in which the verification of beacon outputs during the last several rounds can be batched and verified efficiently. Work~\cite{loe2021practical} on VDF can be investigated and applied to construct a practical VDF-based DRB protocol.
    
    Current DRBs are not Post-Quantum (PQ) secure (except DRB~\cite{li2021postquantum} that does not depend on any third party to construct a quantum-safe beacon). Recent constructions of Post-Quantum VRF~\cite{buser2021post} and Post-Quantum VDF~\cite{PQ-VDF} can be carefully studied and applied to construct practical PQ-secure DRB protocols.

    \item \textit{Using New Threshold Signatures}: Tomescu et al.~\cite{tomescu2020towards} designed a fast BLS-based threshold signature scheme (BLS-TSS). Their scheme has fast signature aggregation and verification. There are some DRBs that use the BLS signature scheme. The new BLS-TSS scheme can be directly applied to these DRBs to improve their performance. Otherwise, a new large-scale, simple DRB protocol can be designed and implemented using this new BLS-TSS scheme.

    \item \textit{Using New Erasure Codes}: All known Verifiable Secret Sharing (VSS) schemes published so far in the open literature (without exception) use the well-known Reed-Solomon codes \cite{reed1960polynomial}. Reed-Solomon codes are Maximum Distance Separable (MDS) erasure codes of type $(t, n)$, where the original message is equally split in $t$ parts and is encoded to $n$ (where $n>t$) parts. In the recent decade, the coding theory community constructed new MDS erasure codes. The most significant line of work was done by Dimakis et al., in~\cite{5550492} where they constructed Minimum Bandwidth Regenerating (MBR) codes (optimal in terms of the repair bandwidth) and Minimum Storage Regenerating (MSR) codes (optimal in terms of the storage). Soon after that, those MDS codes were practically employed in Facebook data centers~\cite{179255}, and new variants of MDS codes e.g.~\cite{ye2017explicit} were proposed. Therefore, it would be interesting to research the potential replacement of the Reed-Solomon code in VSS schemes with another MDS  (MSR) code to improve the performance of VSS-based DRBs.

\end{itemize}

\section{Conclusion}
Within recent years, there has been a dramatic surge in the construction of new Decentralized Randomness Beacon (DRB) protocols due to its emergence in cryptographic protocols. We present the first systematization of knowledge (SoK) for the existing efforts on DRB protocols. This SoK provides a comprehensive review of the design paradigms and approaches of DRB protocols. This SoK can serve as a starting point to explore DRB protocols and can help researchers or practitioners to pick a DRB protocol well-suited for their application.

In this SoK, we presented basic standard definitions of a DRB protocol and its required properties. We discussed the key components and the most significant features of DRB protocols and summarized the existing DRB protocols in Table~\ref{tab:comparison}. We identified several research challenges related to the complexity, scalability, and security of DRB protocols. We highlighted respective solutions to encounter some of the challenges. Finally, we proposed promising research directions for the future design of DRB protocols by employing the new cryptographic components that can help to advance the state-of-the-art of DRB protocols.



\bibliographystyle{splncs04}
\bibliography{reference}

\newpage
\appendix

\section{Secure DRB Protocol} \label{appendix:secure-DRB}
A DRB protocol is said to be secure if for any probabilistic polynomial-time adversary $\mathcal{A}$ corrupting at most $t$ parties in a round $e$, in a security game $\mathcal{G}$ played between the adversary $\mathcal{A}$ and a challenger $\mathcal{C}$, $\mathcal{A}$ has negligible advantage.
\begin{enumerate}
    \item $\mathcal{C}$ executes the setup and sends the public parameters of the system to $\mathcal{A}$.
    \item $\mathcal{A}$ corrupts up to $t$ participants and informs about $t$ corrupted nodes to $\mathcal{C}$.
    \item $\mathcal{C}$ creates the secret and public keys of honest nodes and sends the public keys of honest nodes to $\mathcal{A}$.
    \item $\mathcal{A}$ sends the remaining public parameters (e.g. public keys) of $t$ nodes to $\mathcal{C}$.
    \item $\mathcal{C}$ and $\mathcal{A}$ runs the protocol execution interactively per round where:
    \begin{enumerate}
        \item $\mathcal{C}$ sends all the honest participants' messages to $\mathcal{A}$.
        \item $\mathcal{A}$ decides on the delivery (sends / does not send) of the messages.
        \item At the end of a round $e$, an honest node outputs the protocol transcript. 
    \end{enumerate}
    \item $\mathcal{C}$ samples a bit $b \in \{0,1\}$ and sends either the DRB output based on transcript or a random element.
    \item $\mathcal{A}$ makes a guess $b'$ and the advantage of $\mathcal{A}$ is defined as $|\Pr [b=b'] - \frac{1}{2}|$.
\end{enumerate}


\section{Publicly Verifiable Secret Sharing (PVSS)} \label{appendix:PVSS} 
In a PVSS scheme, a dealer shares a randomly selected secret $s$ among a set of $n$ nodes using an $(n,t+1)$ threshold access-structure. That means, secret $s$ can be recovered from a set of $t+1$ valid shares.
    \begin{definition}(PVSS): \textit{It is defined as a collection of following algorithms:}
    \begin{itemize}[leftmargin=*]
        \item $\mathsf{Setup}(\lambda)$: Given a security parameter $\lambda$, generates the public parameters $pp$ and the public-private key-pair for each node, output the public parameter and public keys $(pp, {pk})$. $pp$ is an implicit input to all the other algorithms.
        \item $\mathsf{Share}(s)$: For a randomly chosen secret $s$, a dealer creates the secret shares for each node $\Vec{S} = (s_1,s_2,\ldots,s_n)$ along with the encryption of the shares $\Vec{E} = (c_1,c_2,\ldots,c_n)$ where $c_i = Enc(s_i)$ and proof of correct encryption $\Vec{\pi} = ({\pi}_1,{\pi}_2,\ldots,{\pi}_n)$. It outputs $(\Vec{S},\Vec{E},\Vec{\pi})$.  
        \item $\mathsf{Verify}(\Vec{E}, \Vec{\pi})$: Given the encrypted shares and the proofs, any external $\mathcal{V}$ can non-interactively verifies if the sharing is correct. It outputs $0$ or $1$.
        \item $\mathsf{Recon}(\Vec{S})$: Given valid set $\Vec{S} \subseteq {\{s_1,s_2,\ldots,s_n\}}^{t+1}$ of  $t+1$ decrypted shares, it reconstructs the secret and outputs $s$. 
    \end{itemize}
    
    \end{definition}
    
    
        
        


 \begin{figure*}[th]
    \centering
    \includegraphics[width = \textwidth]{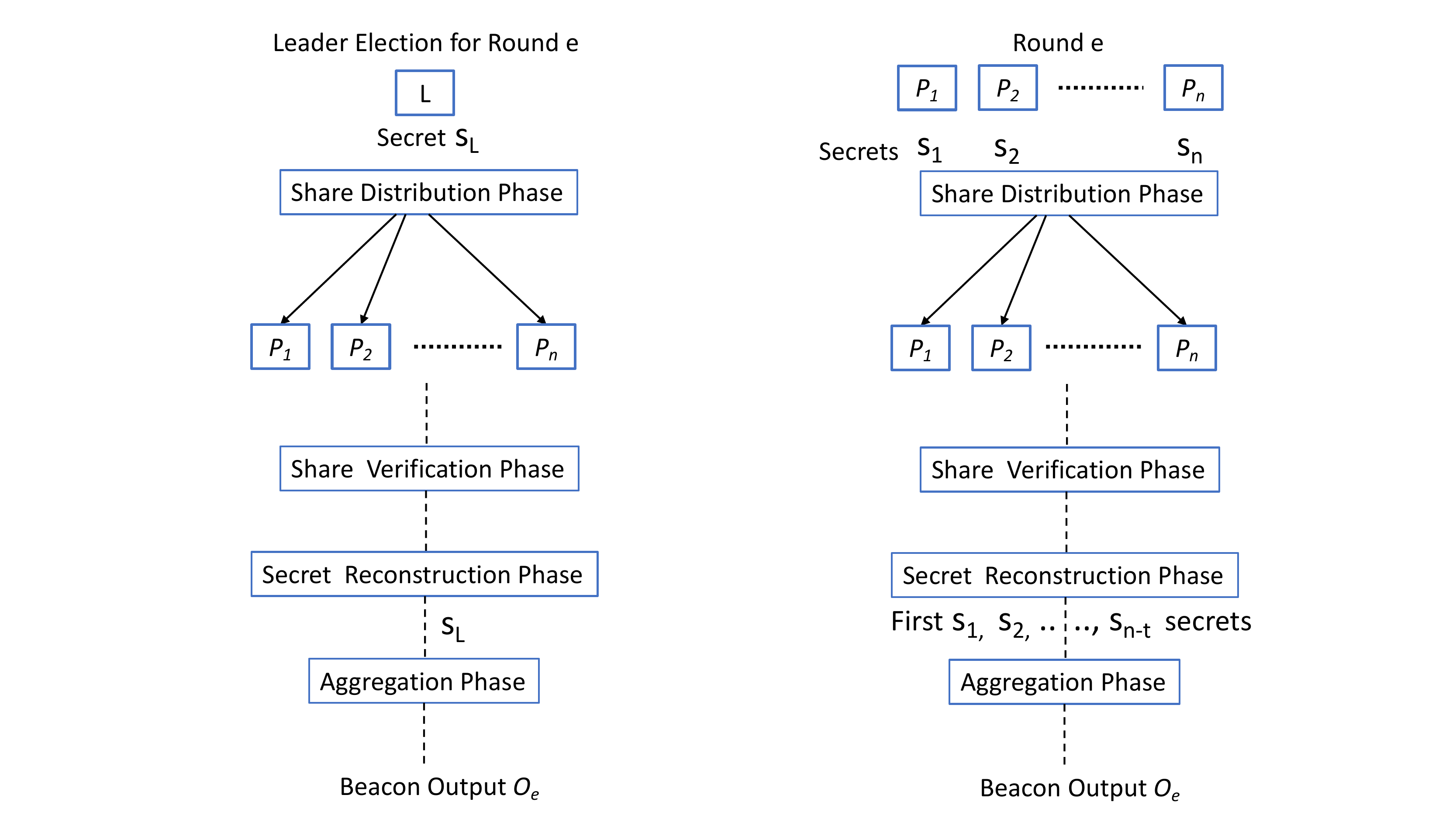}
     \vspace{-0.7cm}
    \caption{PVSS-based DRB protocols with and without leader}
    \vspace{-0.5cm}
    \label{fig:I-DRB}
\end{figure*}
 
 In a DRB protocol involving a leader, once the setup phase is completed, for the round $e$, first a leader election algorithm $\mathsf{LeaderElec}(e,O_{e-1},P_1, P_2, \ldots, P_n)$ is executed and a leader $L_e$ is selected. The election algorithm can be \textit{round-robin selection} or \textit{sampling uniformly at random}. The leader $L_e$ chooses a secret value $s_{L_e}$ (either a new value or previously committed value in the previous round) and executes the PVSS scheme for secret $s_{L_e}$. At the end of round $e$, DRB output $O_e$ is generated using the reconstructed secret and the previous round (or rounds') output value. Figure~\ref{fig:I-DRB} depicts leader and non-leader-based DRB protocols. In the first sub-figure, a leader is elected, followed by leader's secret is shared and beacon output is produced. In the second sub-figure, all participants randomly choose secrets at the start of the round and further share the encrypted shares of the secret to all the other participants. In the final stage, the first n-t reconstructed (or decrypted) shares are used to obtain beacon output.

 \begin{definition}(PVSS-based Interactive Decentralized Randomness Beacon (I-DRB)) Given a set of participants $\mathcal{P} = (P_1, P_2, \ldots, P_n)$, a PVSS-based I-DRB without leader can be defined as a tuple $\mathcal{B}$ of polynomial-time algorithms:\\
$\mathcal{B} = (\mathsf{Setup, Share, Verify, Recon, Aggregation})$
\begin{itemize}[leftmargin=*]
    \item $\mathsf{Setup}(e,\lambda)$: Set the round $e = 1$. Run $\mathsf{PVSS.Setup}(\lambda)$ and generate public parameter $pp$ and key-pairs $(sk_i,pk_i)$ for each participant. 
    \item $\mathsf{Share}(e)$: For a round $e$, each participant $P_i$ runs $\mathsf{PVSS.Share}(s_i)$ for a randomly chosen value $s_i$ from the input space and gets $(\Vec{S}_i,\Vec{E}_i,\Vec{\pi}_i)$. $P_i$ shares the encrypted shares and corresponding proofs $(\Vec{E}_i,\Vec{\pi}_i)$  with other participants.
    \item $\mathsf{Verify}(e,\{\Vec{E},\Vec{\pi}\})$: Each party $P_j$ runs the share verification algorithm $\mathsf{PVSS.Verify}\\(\Vec{E}_i,\Vec{\pi}_i); \forall i, i\neq j $ on every shared secret. Let $\mathcal{C}$ be the set of first $n-t$ participants who have correctly shared their random secret values.
    \item $\mathsf{Recon}(e,\{\Vec{S}_i\})$: Each party $P_i$ in $\mathsf{C}$ opens the Shamir secret $s_i$ and the randomness used, other participants $P_j;  \forall j, j\neq i$ verify if it is consistent with sharing posted during $\mathsf{Share}$ phase. If a party $P_i$ refuses to open its secret $s_i$, the secret is reconstructed by executing $\mathsf{PVSS.Recon}(\Vec{S}_i)$. 
    \item $\mathsf{Aggregation}(e,\{s_i\})$: Once the valid decrypted or reconstructed shares are available for the parties $P_i \in \mathcal{C}$. The beacon output is generated by executing a function $f$ on input a set of valid shares $\{s_i\}$. This function $f$ takes  all the valid shares $\{s_i\}$ (additionally previous beacon outputs) as input and aggregates these input values to generate the beacon output $O_e$ for round $e$.
\end{itemize}
\end{definition}

\section{Verifiable Delay Function (VDF)} \label{appendix:VDF}
Verifiable delay function $f : \mathcal{X} \rightarrow \mathcal{Y}$ was defined formally by Boneh et al.~\cite{Boneh2018}. After the introduction of VDF, two new proposals~\cite{Wesolowski2019}~\cite{pietrzak:LIPIcs:2018} were presented. A VDF has properties of \textit{Sequentiality}, \textit{Uniqueness} and \textit{$\epsilon$-Evaluation time}.

\begin{definition}(VDF): \textit{A VDF is defined as a tuple of following algorithms:}
\begin{itemize}[leftmargin=*]
    \item $\mathsf{Setup}(\lambda, T$): It is a randomized algorithm that takes security parameter $\lambda$, time parameter $T$ and outputs public parameter $pp := (\mathbb{G}, N, H, T)$, where $\mathbb{G}$ is a finite abelian group of unknown order, $N$ is an RSA modulus, and $H: \mathcal{X} \rightarrow \mathbb{G}$ is a hash function.
    \item $\mathsf{Eval}({pp}, x, T$): The evaluation algorithm  applies
    $T$ squarings in $\mathbb{G}$ starting with $H(x)$ and
    outputs the value $y\leftarrow H(x)^{\left ( 2^{T} \right )} \mathsf{mod}\:N$, along with a proof $\pi$.
    \item $\mathsf{Verify}({pp}, x, y, \pi, T$): The verification algorithm outputs a bit $\in \{0,1\}$, given the input as public parameter $\mathsf{pp}$, input value $x$, output value $y$, proof $\pi$, and time parameter $T$.
\end{itemize}
\end{definition}
 

\section{Verifiable Random Function (VRF)} \label{appendix:VRF}
VRF has properties of Uniqueness, Collision resistance and Pseudorandomness.
\begin{definition}(VRF): \textit{A VRF is defined as a tuple of following algorithms:} 
\begin{itemize}[leftmargin=*]
    \item $\mathsf{KeyGen}(r)$: On input value $r$, the algorithm generates a secret key $sk$ and a verification key $vk$. 
    \item $\mathsf{Eval}(sk, M)$:  Evaluation algorithm produces pseudorandom output $O$ and the corresponding proof $\pi$ on input sk and a  message $M$.
    \item $\mathsf{Verify}(vk, M, O, \pi)$: Verify algorithm outputs 1 if and only if the output produced by evaluation algorithm is $O$ and it is verified by the proof $\pi$ given the verification key $vk$ and the message $M$.
\end{itemize}
\end{definition}


\section{Homomorphic Encryption (HE)} \label{appendix:HE}
    \begin{definition}(HE): \textit{An HE scheme is defined as a set of following alogorithms:}\
    \begin{itemize}[leftmargin=*]
        \item $\mathsf{Setup}(1^{\lambda})$: Given security parameter $\lambda$, Output global parameters $param$s.
        \item $\mathsf{KeyGen}(params) $: Given global parameters $param$, output a public-private key-pair $(pk,sk)$. 
        \item $\mathsf{Enc}(params, pk,\mu)$: Given a message $\mu \in R_\mathcal{M}$, output a ciphertext $c$. 
        \item $\mathsf{Dec}(params, sk, c)$: Given a ciphertext $c$, output a message $\mu^* \in R_\mathcal{M}$.
        \item $\mathsf{Eval}(pk, f , c_1 , . . . , c_l )$: Given the inputs as public key $pk$, a function $f : R_\mathcal{M}^l \rightarrow R_\mathcal{M}$ which is an arithmetic circuit over $R_\mathcal{M}$, and a set of $l$ ciphertexts $c_1 , . . . , c_l $, output a ciphertext $c_f$. 
    \end{itemize}
    
    \end{definition}
    
    In the above scheme, the message space $\mathcal{M}$ of the encryption schemes is a ring $R_\mathcal{M}$, and the functions to be evaluated are represented as arithmetic circuits over this ring, composed of addition and multiplication gates. HE can be categorized into: \textit{Partially HE} that supports only addition or multiplication; \textit{Somewhat HE} that allows both operations but with limited times; \textit{Fully HE} that supports arbitrary computation by allowing both operations with unlimited times.

    



\section{Hybrid DRB Protocols} \label{appendix:other-DRBs}
There are many more DRB protocols. Some of these protocols use more than one crypto primitive  to achieve all DRB properties with better efficiency/optimization.

\begin{itemize}[leftmargin=*]
    \item Mt. Random (PVSS + T(VRF)){(eprint 2021/1096)}: It is a multi-tiered DRB protocol that combines PVSS, VRF, and Threshold VRF (TVRF) to construct a DRB with optimal efficiency and without compromising security guarantees of DRB. It is a flexible architecture for DRB where each tier runs a separate beacon based on PVSS, VRF, and TVRF, and output of one tier works as a seed for the next tier. Being constructed using different crypto-primitives, each tier differs in the provided randomness and complexity. Due to that, a high-level application can decide on which tier to use to obtain randomness.
    \item Harmony (VRF + VDF){(https://harmony.one/whitepaper.pdf)}: Harmony is a sharding-based, provably secure, and scalable blockchain. In Harmony, nodes compute local entropy by executing VRF using their secret keys. DRB output is computed using VDF where the input for the VDF is constructed from a threshold number of VRF evaluations from pairwise different nodes. DRB output is made pseudorandom by applying a random oracle on VDF output.
    
    \item CRAFT (TLP + VDF){(eprint 2020/784)}: Baum et. al first construct UC-secure publicly verifiable TLP and UC-secure VDF. To construct DRB, they replace the commitments with the UC-secure TLP in the standard commit-reveal coin-tossing protocol. Their construction achieves $\mathsf{O}(n)$ communication to generate DRB output. DRB output can be obtained as fast as the communication channel delay allows when nodes communicate their TLPs faster.
    \item VeeDo (STARK+VDF){(https://github.com/starkware-libs/veedo)}: It is based on STARK-based VDF. STARK is a post-quantum secure zero-knowledge proof protocol. VeeDo is a smart-contract-based DRB where a beacon smart contract and a verifier smart contract is placed on-chain. However, heavy computational parts involving VDF and STARK prover are kept off-chain. A VDF is run on a seed $s$ from a block hash to compute the DRB output and a proof is computed using the STARK prover. The VDF output and the proof are sent to the on-chain contracts for verification and subsequently publishing.
    \item STROBE (RSA-based){(eprint 2021/1643)}: It is a history-generating DRB (HGDRB). It allows efficient generation of previous beacon outputs given only the current beacon value and public key. It is based on origin squaring based RSA approach of Beaver. It is well-suited for practical applications especially in streaming designs where it allows client software to generate game states by computing every missing beacon value and state. It is NIZK-free, concisely self-verifying and can be efficiently used in blockchain and voting systems. 
    \item OptRand (Bilinear paring-based PVSS + NIZK){(eprint 2022/193)}: It is an optimally responsive DRB protocol. It employs a pairing-based PVSS scheme together with a NIZK proof system to produce DRB outputs. Despite the synchrony of the network, it can provide an optimal response and can progress. Therefore, OptRand can provide availability at actual network speed during optimistic conditions. It is reconfiguration-friendly and has low communication complexity and low latency while generating beacon outputs.
\end{itemize}

\end{document}